\documentclass[12pt]{article}
\usepackage{latexsym,epsfig,graphicx,a4wide,amssymb}

\def\de{\partial}
\def\a{\alpha}
\def\b{\beta}
\def\g{\gamma}
\def\G{\Gamma}
\def\d{\delta}

\def\f{\phi}{\rm }
\def\la{\lambda}
\def\La{\Lambda}
\def\k{\kappa}
\def\m{\mu}
\def\n{\nu}
\def\r{\rho}

\def\s{\sigma}

\def\th{\theta}
\def\z{\zeta}
\def\x{\chi}
\def\be{\begin{equation}}
 \def\ee{\end{equation}}
 \def\bea{\begin{eqnarray}}
 \def\eea{\end{eqnarray}}
 \def\a{\alpha}
 \def\b{\beta}
 \def\g{\gamma}
 \def\d{\delta}
 
 \def\s{\sigma}
\def\G{\Gamma}
\def\L{\Lambda}

\newcommand{\fr}{\frac}

\def\2{\frac{1}{2}}
\def\4{\frac{1}{4}}

\catcode`\@=11

\@addtoreset{equation}{section}

\def\@normalsize{\@setsize\normalsize{15pt}\xiipt\@xiipt
\abovedisplayskip 14pt plus3pt minus3pt%
\belowdisplayskip \abovedisplayskip
\abovedisplayshortskip  \z@ plus3pt%
\belowdisplayshortskip  7pt plus3.5pt minus0pt}
\def\small{\@setsize\small{13.6pt}\xipt\@xipt
\abovedisplayskip 13pt plus3pt minus3pt%
\belowdisplayskip \abovedisplayskip
\abovedisplayshortskip  \z@ plus3pt%
\belowdisplayshortskip  7pt plus3.5pt minus0pt
\def\@listi{\parsep 4.5pt plus 2pt minus 1pt
            \itemsep \parsep
            \topsep 9pt plus 3pt minus 3pt}}

\def\underline#1{\relax\ifmmode\@@underline#1\else
        $\@@underline{\hbox{#1}}$\relax\fi}
\@twosidetrue \relax

\catcode`@=12

\catcode`\@=11

\def\section{\@startsection{section}{1}{\z@}{3.5ex plus 1ex minus
   .2ex}{2.3ex plus .2ex}{\large\bf}}

\def\ps@headings{\def\@oddfoot{}\def\@evenfoot{}
\def\@oddhead{\hbox{}\hfill
        \makebox[.5\textwidth]{\raggedright\ignorespaces --\thepage{}--
        \hfill }}
\def\@evenhead{\@oddhead}
\def\subsectionmark##1{\markboth{##1}{}}
}

\ps@headings

\catcode`\@=12

\begin{document}

\begin{titlepage}
%
%


%

\begin{centering}
\vspace{1cm}
{\Large {\bf Black Holes on Thin 3-branes of  Codimension-2\\ and their Extension  into the Bulk}}\\

\vspace{1.5cm}

 {\bf Bertha Cuadros-Melgar $^{*}$} \\
 \vspace{.2in}
Departamento de F\'isica, Universidad de Santiago de Chile,\\
Casilla 307, Santiago, Chile\\
 \vspace{.2in}
 {\bf Eleftherios Papantonopoulos}$^{**}$, {\bf Minas~Tsoukalas} $^{\flat}$ \\{\bf and}\\
  {\bf Vassilios Zamarias}$^{\natural}$\\
\vspace{.2in}

 Department of Physics, National Technical University of
Athens, \\
Zografou Campus GR 157 73, Athens, Greece. \\
\vspace{3mm}

\end{centering}
\vspace{2cm}

\begin{abstract}

We discuss black hole solutions in six-dimensional gravity with a
Gauss-Bonnet term in the bulk and an induced gravity term on a
 thin  3-brane of codimension-2. We show that these
black holes can be localized on the brane, and they can further be
extended into the bulk by a warp function. These solutions have
regular horizons and no other curvature singularities appear apart
from the string-like ones. The projection of the Gauss-Bonnet term
on the brane imposes a constraint relation  which requires the
presence of matter in the extra  dimensions.

\end{abstract}

\vspace{2.5cm}
\begin{flushleft}
$^{*}~~$ e-mail address: berthaki@gmail.com \\
$^{**} ~$ e-mail address: lpapa@central.ntua.gr \\
$ ^{\flat}~~$ e-mail address: minasts@central.ntua.gr\\
$ ^{\natural}~~$ e-mail address: zamarias@central.ntua.gr

\end{flushleft}
\end{titlepage}

\section{Introduction}

Recently there has been a growing interest in codimension-2
braneworlds. The most attractive feature of these models is that
the vacuum  energy (tension) of the brane
 instead of curving the brane world-volume, merely induces a deficit angle in the
 bulk solution around the brane \cite{Chen:2000at}. This observation
led  several people to utilize this property in order to self-tune
the effective cosmological constant  to zero and provide a
solution to the cosmological constant problem \cite{6d}. However,
soon it was realized \cite{Cline}
   that one can only find nonsingular
solutions if the brane energy momentum tensor is proportional to
its induced metric. To reproduce an effective four-dimensional
Einstein equation on the brane one has to introduce a cut-off
(brane thickness) \cite{Kanno:2004nr,Vinet:2004bk,Navarro:2004di}
with the price of loosing the predictability of the theory.
Alternatively, in the thin brane limit four dimensional gravity
is recovered as the dynamics of the induced metric on the brane if
the gravitational action is modified by the inclusion of either a
Gauss-Bonnet term~\cite{Bostock:2003cv} or an induced gravity term
on the brane~\cite{Papantonopoulos:2005ma}.

We are still lacking an understanding of time dependent
cosmological solutions in codimension-2 braneworlds. In the thin
brane limit, because the energy momentum tensors on the brane and
in the bulk are related, the brane equation of state and energy
density are tuned and we cannot get the standard cosmology on the
brane~\cite{Kofinas:2005py,Papantonopoulos:2005nw}. One then has
to regularize the codimension-2 branes by introducing some
thickness and then consider matter on them
\cite{regular,PST,ppz,tas}. To have a cosmological evolution on
the regularized branes the brane world-volume should be expanding
and in general
 the bulk space should also evolve in time. This is a formidable task,
  so an alternatively approach was followed in~\cite{Papantonopoulos:2007fk,Minamitsuji:2007fx}
by considering a codimension-1 brane moving in the regularized
static background. The resulting cosmology, however, was unrealistic
having a negative Newton's constant (for a review on the cosmology
in six dimensions see~\cite{Papantonopoulos:2006uj}).

We do not either fully understand black hole solutions on
codimension-2 braneworlds. Recently a six-dimensional black hole
localized on a 3-brane of codimension-2~\cite{Kaloper:2006ek} was
proposed. These solutions are generalization of the $4D$ Aryal,
Ford, Vilenkin~\cite{Aryal:1986sz,Achucarro:1995nu} black hole
pierced by a cosmic string adjusted to the codimension-2 branes
with a conical structure in the bulk and deformations
 accommodating the deficit angle. However, it is not
clear how to realize these solutions in the thin brane limit where
high curvature terms are needed to accommodate matter on the
brane. Generalizations to include rotations were presented
in~\cite{Kiley:2007wb} and perturbative analysis of these black
holes were carried out in~\cite{alBinni:2007gk,Chen:2007ay}.

The  localization of a black hole on the brane and its extension
to the bulk is a difficult task. In codimension-1 braneworlds the
first attempt was to consider the Schwarzschild metric and study
its black string extension into the bulk~\cite{Chamblin:1999by}.
Unfortunately, as suspected by the authors, this string is
unstable to classical linear perturbations~\cite{BSINS} (for a
recent review see~\cite{Harmark:2007md}). Since then, several
authors have attempted to find the full metric using numerical
techniques \cite{BHNUM}. Analytically, the brane metric equations
of motion were considered with the only bulk input coming from
the projection of the Weyl tensor~\cite{SMS} onto the brane.
Since this system is not closed because it contains an unknown
bulk dependent term, assumptions have to be made either in the
form of the metric or in the Weyl term~\cite{BBH}. So far there is no
clear evidence of what the brane black hole metric is, however,
some interesting features which do occur are wormholes and
singular horizons \cite{GWBD,IFeat}. Analysis of the stability and
thermodynamics of these solutions were worked out in \cite{ACMPM}.

 A lower dimensional version of a black hole living
on a (2+1)-dimensional braneworld was considered in~\cite{EHM} by
Emparan, Horowitz, and Myers. They based their analysis on the
so-called C-metric~\cite{Kinnersley:zw} modified by a cosmological
constant term. They found a BTZ black hole~\cite{Banados:1992wn}
on the brane which can be extended as a BTZ black string in a
four-dimensional AdS bulk. Their thermodynamical stability
analysis showed that the black string remains a stable
configuration when its transverse size is comparable to the
four-dimensional AdS radius, being destabilized by the
Gregory-Laflamme instability~\cite{BSINS} above that scale,
breaking up to a BTZ black hole on a 2-brane.

Three-dimensional gravity, because of its simplicity, is widely
recognized as a useful laboratory to study important issues of
general relativity. Earlier work on (2+1)-gravity~\cite{DJtH, DJ}
has been followed by many authors  studying various aspects of
classical and quantum gravity~(for a review see
~\cite{Carlip:1995zj}). In spite of the fact that
 general relativity in
(2+1) dimensions has neither Newtonian limit  nor propagating
degrees of freedom, a black hole solution was found (BTZ black
hole~\cite{Banados:1992wn}). The BTZ black hole differs from the
Schwarzschild and Kerr solutions in some important aspects: it
has a conical-like axially symmetric metric, it is asymptotically
anti-de Sitter rather than asymptotically flat, and it has no
curvature singularity at the origin. Nonetheless, it is clearly a
black hole: it has an event horizon and (in the rotating case) an
inner horizon, it appears as the final state of collapsing matter,
and it has thermodynamic properties much like those of a
(3+1)-dimensional black hole. A singular solution at the origin
was presented in~\cite{Zanelli1996} as a result of the coupling of
BTZ black hole to a conformal matter field, and it was further
extended in~\cite{Henneaux:2002wm}.

In our previous work~\cite{CuadrosMelgar:2007jx} we studied black
holes on an infinitely thin conical 2-brane and their extension
into a five-dimensional bulk with a Gauss-Bonnet term. We had
found two classes of solutions. The first class consists of the
familiar BTZ black hole which solves the junction conditions on a
conical 2-brane in vacuum. These solutions in the bulk are BTZ
string-like objects with regular horizons and no pathologies. The
warping to five-dimensions depends on the length $\sqrt{\alpha}$
where $\alpha$ is the Gauss-Bonnet coupling, and this length scale
defines the shape of the horizon. Consistency of the bulk
solutions requires a fine-tuned relation between the Gauss-Bonnet
coupling and the five-dimensional cosmological constant. The
second class of solutions consists of BTZ black holes with short
distance corrections. These solutions correspond to a BTZ black
hole conformally dressed with a scalar
field~\cite{Zanelli1996,Henneaux:2002wm}. Localization of these
black holes on the 2-brane leads to the interesting result that
the energy-momentum tensor required to support such solutions on
the brane corresponds to the energy-momentum tensor of a scalar
field in the
 limit $r/L_{3}<<1$, where $L_{3}$ is the length scale of the
three-dimensional AdS space and $r$ the radial distance on the
brane. Also these solutions have black string-like extensions into
the bulk.

In this work we generalize  our previous work to black objects in
six-dimensional brane-worlds of codimension-2. We find solutions
of four-dimensional Schwarzschild-AdS black holes on the brane
which in the six-dimensional spacetime look like black string-like
objects with regular horizons. The warping to extra dimensions
depends on the Gauss-Bonnet coupling which is fine-tuned to the
six-dimensional cosmological constant. In the case of constant
deficit angle the localization of the four-dimensional black hole
requires matter in the two extra dimensions. The energy-momentum
tensor corresponding to this matter scales as $1/r^6$.  This fact
defines a length scale in the six-dimensional spacetime above
which we recover the standard four-dimensional General Relativity
(GR), while at small distances GR is strongly modified. There are
also solutions with variable deficit angle, in which case matter
is also necessary in the other directions. However, consistency of
the bulk equations requires the deficit angle to be constant.

The presence of the Gauss-Bonnet term in codimension-2 braneworlds
has important consequences in our solutions. Its projection on the
brane gives a consistency relation~\cite{Papantonopoulos:2005ma}
that dictates the form of the solutions. It allows black string
solutions in five dimensions and in six dimensions it specifies
the kind of matter which is needed  in the bulk in order to
support a black hole solution on the brane.

The paper is organized as follows. In section 2 we present in a
self-contained  way the BTZ string-like  solutions of the
five-dimensional case. In section 3 we discuss the black
string-like solutions of the six-dimensional Einstein equations
for constant and variable deficit angles. To complete our
solutions we introduce branes, and solving the junction equations
we  find the conditions to localize the black holes
on these branes. In section 4 we discuss the special r$\hat o$le
played by the Gauss-Bonnet term, and finally, in section 5 we
conclude.

\section{BTZ String-Like Solutions in Five-Dimensional Braneworlds of Codimension-2}

We consider the following gravitational action in five dimensions
with a Gauss-Bonnet term in the bulk and an induced
three-dimensional curvature term on the brane
\begin{eqnarray}\label{AcGBIG}
S_{\rm grav}&=&\frac{M^{3}_{5}}{2}\left\{ \int d^5 x\sqrt{-
g^{(5)}}\left[ R^{(5)}
+\alpha\left( R^{(5)2}-4 R^{(5)}_{MN}R^{(5)MN}+
R^{(5)}_{MNKL}R^{(5)MNKL}\right)\right] \right.\nonumber\\
&+& \left. r^{2}_{c} \int d^3x\sqrt{- g^{(3)}}\,R^{(3)}
\right\}+\int d^5 x \mathcal{L}_{bulk}+\int d^3 x
\mathcal{L}_{brane}\,,\label{5daction}
\end{eqnarray}
where $\alpha\, (\geq0)$ is the GB coupling constant and
$r_c=M_{3}/M_{5}^3$ is the induced gravity ``cross-over" scale
(marking the transition from 3D to 5D gravity).
In the above action $M_{5}$ is the five-dimensional Planck mass
and  $M_{3}$ is the three-dimensional one.
 The above induced term has been
written in the particular coordinate system in which the metric is
\be ds_5^2=g_{\m\n}(x,\rho)dx^\m
dx^\n+a^{2}(x,\rho)d\rho^2+L^2(x,\rho)d\th^2~,\label{5dmetric} \ee
where $g_{\mu\nu}(x,0)$ is the braneworld metric and $x^{\mu}$
denote three  dimensions, $\mu=0,1,2$~ whereas $\rho,\th$ denote
the radial and angular coordinates of the two extra dimensions
(the $\rho$ direction may
 or may not  be compact and the $\th$ coordinate ranges form $0$ to $2\pi$).
Capital $M$,~$N$ indices will take values in the five-dimensional
space. Note that we have assumed that there exists an azimuthal
symmetry in the system, so that both the induced three-dimensional
metric and the functions $a$ and $L$ do not depend on $\th$.

The Einstein equations resulting from the variation of the
action~(\ref{5daction}) are \be
 G^{(5)N}_M + r_c^2
G^{(3)\n}_\m g_M^\m g^N_\n {\d(\rho) \over 2 \pi L}-\alpha
H_{M}^{N} =\frac{1}{M^{3}_{5}} \left[T^{(B)N}_M+T^{(br)\n}_\m
g_M^\m g^N_\n {\d(\rho) \over 2 \pi L}\right]~, \label{einsequat3}
\ee where \bea
 H_M^N&=& \left[{1
\over 2}g_M^N (R^{(5)~2}
-4R^{(5)~2}_{KL}+R^{(5)~2}_{ABKL})\right.-2R^{(5)}R^{(5)N}_{M}\nonumber\\
&&+4R^{(5)}_{MP}R^{NP}_{(5)}\phantom{{1 \over 2}}~\left.
+4R^{(5)~~~N}_{KMP}R_{(5)}^{KP} -2R^{(5)}_{MKL P}R_{(5)}^{NKL
P}\right]~. \label{gaussbonnet} \eea To obtain the braneworld
equations we expand the metric around the brane as \be
L(x,\rho)=\beta(x)\rho+O(\rho^{2})~. \ee At the boundary of the
internal two-dimensional space where the 2-brane is situated the
function $L$ behaves as $L^{\prime}(x,0)=\beta(x)$, where a prime
denotes derivative with respect to $\rho$. We also demand that the
space in the vicinity of the conical singularity is regular which
imposes the supplementary conditions that $\de_\m \b=0$ and
$\partial_{\rho}g_{\mu\nu}(x,0)=0$~\cite{Bostock:2003cv}.

The extrinsic curvature in the particular gauge $g_{\rho \rho}=1$
that we are considering  is given by $K_{\m\n}=g'_{\m\n}$.
 The above decomposition will be helpful in the following for finding the induced dynamics on the brane.
We will now use the fact that the second derivatives of the metric
functions contain $\d$-function singularities at the position of
the brane. The nature of the singularity then gives the following
relations \cite{Bostock:2003cv} \bea
{L'' \over L}&=&-(1-L'){\d(\rho) \over L}+ {\rm non-singular~terms}~,\\
{K'_{\m\n} \over L}&=&K_{\m\n}{\d(\rho) \over L}+ {\rm
non-singular~terms}~. \eea

From the above singularity expressions and using the Gauss-Codazzi
equations, we can  match the singular parts of the Einstein
equations (\ref{einsequat3}) and get the following ``boundary"
Einstein equations
 \be G^{(3)}_{\m\n}={1 \over M_{(5)}^3 (r_c^2+8\pi
(1-\b)\a)}T^{(br)}_{\m\n}+{2\pi (1-\b) \over r_c^2+8\pi
(1-\b)\a}g_{\m\n} \label{einsteincomb3}~. \ee

Note that in the above boundary Einstein equations, as a result of
the Gauss-Codazzi reduction procedure, there will also appear
terms proportional to the extrinsic curvature and terms coming
from the GB term in the bulk. However, if we allow only conical
singularities there is no contribution from these
terms~\cite{Bostock:2003cv} (see next section for the most general
case). Also observe, that the presence of the induced gravity on
the brane or the GB term in the bulk is necessary in order to have
a non zero energy momentum tensor on the brane.

We assume that there is a localized (2+1) black hole on the brane.
The brane metric is \be
ds_{3}^{2}=\left(-n(r)^{2}dt^{2}+n(r)^{-2}dr^{2}+r^{2}d\phi^{2}\right)~,
\label{3dmetric}\ee where $0\leq r< \infty$ is the radial
coordinate, and $\phi$ has
 the usual periodicity $(0,2\pi)$.
 We will look for black string solutions of the
 Einstein equations~(\ref{einsequat3}) using the
 five-dimensional metric~(\ref{5dmetric}) in the form
\be ds_5^2=f^{2}(\rho)\left(-n(r)^{2}dt^{2}+n(r)^{-2}dr^{2}+r^{2}
d\phi^{2}\right)+a^{2}(r,\rho)d\rho^2+L^2(r,\rho)d\th^2~.\label{5smetricc}
\ee

 The space outside the
 conical singularity is regular, therefore, we  demand that the warp function $ f(\rho) $ is
  also regular
 everywhere. We assume that there is only a cosmological constant $\Lambda_{5}$
in the bulk and we take $a(r,\rho)=1$. Then, from the bulk Einstein
equations \be G^{(5)}_{MN}-\alpha
H_{MN}=-\frac{\Lambda_{5}}{M^{3}_{5}}g_{MN}~,\ee combining the
$(rr,\phi \phi)$ equations we get

 \be
 \left(\dot{n}^{2}+n \ddot{n}-\frac{n \dot{n}}{r}\right)\left(1-4\alpha \frac{L''}{L}\right)=0~,\label{173}
 \ee
 while a combination of the $(\rho \rho, \theta \theta)$ equations
 gives
 \be
 \left(f''-\frac{f'L'}{L}\right)\left[3-4\frac{\alpha}{f^{2}}\left(\dot{n}^{2}+n
  \ddot{n}+2\frac{n \dot{n}}{r}+3f'^{2}
 \right)\right]=0\label{183}~,
 \ee
 where a  dot denotes derivatives with respect to $r$. The
 solutions of the equations (\ref{173}) and (\ref{183})
are summarized in the following table~\cite{CuadrosMelgar:2007jx}\\

\begin{table}[here]
\begin{center}
\begin{tabular}{|c|c|c|c|c|c|}
  \hline
  $n(r)$ & $f(\r)$ & $L(\r)$ & $-\L_5$ & Constraints \\
  \hline
  BTZ & $\cosh\left(\frac{\r}{2\,\sqrt{\a}}\right)$ & $\forall L(\r)$ &
$\frac{3}{4\a}$ &
$L_3^2=4\,\a$ \\
  BTZ & $\cosh\left(\frac{\r}{2\,\sqrt{\a}}\right)$ & $2\,\b\,\sqrt{\a}\,\sinh\left(\frac{\r}{2\,\sqrt{\a}}\right)$ &
  $\frac{3}{4\a}$ & - \\
  BTZ & $\cosh\left(\frac{\r}{2\,\sqrt{\a}}\right)$ & $2\,\b\,\sqrt{\a}\,\sinh\left(\frac{\r}{2\,\sqrt{\a}}\right)$ &
  $\frac{3}{4\a}$ & $L_3^2=4\,\a$ \\
  BTZ & $\pm 1$ & $\frac{1}{\g}\,\sinh\left(\g\,\r\right)$ & $\frac{3}{l^2}$ & $\g=\sqrt{-\frac{2\L_5}{3+4\a\L_5}}$ \\
  $\forall n(r)$ & $\cosh\left(\frac{\r}{2\,\sqrt{\a}}\right)$ & $2\,\b\,\sqrt{\a}\,\sinh\left(\frac{\r}{2\,\sqrt{\a}}\right)$ &
  $\frac{3}{4\a}$ & -  \\
  $\sqrt{-M+\frac{r^2}{L_3^2}-\frac{\z}{r}}$ & $\cosh\left(\frac{\r}{2\,\sqrt{\a}}\right)$ &
$2\,\b\,\sqrt{\a}\,\sinh\left(\frac{\r}{2\,\sqrt{\a}}\right)$ &
$\frac{3}{4\a}$ &
$L_3^2=4\,\a$ \\
  $\sqrt{-M+\frac{r^2}{L_3^2}-\frac{\z}{r}}$ & $\pm 1$ & $2\,\b\,\sqrt{\a}\,\sinh\left(\frac{\r}{2\,\sqrt{\a}}\right)$
  & $\frac{1}{4\a}$ & $\L_5=-\frac{1}{4\a}=-\frac{3}{L_3^2}$ \\
  \hline
\end{tabular}
\end{center}
\caption{BTZ String-Like Solutions in Five-Dimensional Braneworlds
of Codimension-2}\label{table1}
\end{table}

In the above table $L_3$ is the length scale of $AdS_{3}$ space.
Note that in all solutions there is a fine-tuned relation between
the Gauss-Bonnet coupling $\alpha$ and the five-dimensional
cosmological constant $\Lambda_{5}$, except for the solution in
the fourth row. Also observe that the  solution in the third row
is a kind of combination of the solutions in the first and second
row. This is a result of the way we  solve the factorized
equations (\ref{173}) and (\ref{183})~\cite{CuadrosMelgar:2007jx}.

To introduce a brane we must solve the corresponding junction
conditions given by the Einstein equations on the brane
(\ref{einsteincomb3}) using the induced metric on the brane given
by (\ref{3dmetric}). For the case when $n(r)$ corresponds to the
BTZ black hole $n^{2}(r)=-M+\frac{r^{2}}{L_3^{2}}$, and the brane
cosmological constant is given by $\Lambda_{3}=-1/L_3^{2}$, we
found that the energy-momentum tensor is null. Therefore, the BTZ
black hole is localized on the brane in vacuum.

When $n(r)$ is of the form given by \be
n(r)=\sqrt{-M+\frac{r^2}{L_3^2}-\frac{\z}{r}}~,
\label{nsolu5final} \ee which is the BTZ black hole solution with
a short distance correction term, we can go back to
(\ref{einsteincomb3}) and solve for $T^{br}_{\mu\nu}$. Then we
find  that the matter source necessary to sustain such a solution
on the brane is given by
 \be T_\alpha ^\beta
=  \hbox{diag } \left(
\frac{\zeta}{2r^3},\frac{\zeta}{2r^3},-\frac{\zeta}{r^3} \right)\,
,\label{braneEnerMom} \ee which is conserved on the
brane~\cite{Kofinas:2005a}. Interesting enough, for a scalar field
conformally coupled to BTZ~\cite{Zanelli1996,Henneaux:2002wm}, the
energy-momentum tensor needed to support such a solution at a
certain limit reduces to (\ref{braneEnerMom}) which is necessary
to localize this black hole on the conical 2-brane.

These solutions extend the brane BTZ black hole into the
bulk. Calculating the square of the Riemann tensor we find that at
the AdS horizon ($\rho \rightarrow \infty$) all solutions give
finite result and hence the only singularity is
  the  BTZ-corrected black hole singularity extended into the bulk.
  The warp function $f^{2}(\rho)$ gives the shape of a
'throat' to
    the horizon
of the BTZ string-like solution. The size of the horizon is
defined by the scale $\sqrt{\alpha}$ and this scale is fine-tuned
to the length scale of the five-dimensional AdS space.

\section{Black String-Like solutions in Six-Dimensional Braneworlds of Codimension-2 }

In this section we will look for black string solutions  in
six-dimensions with conical singularities. We consider the
gravitational action (\ref{AcGBIG}) in six dimensions
\begin{eqnarray}\label{act6}
S_{grav}&=& \frac{M_{6}^4}{2} \left\{ \int d^6 x \sqrt{-g^{(6)}} \left[ R^{(6)} + \alpha \left( R^{(6)2} - 4 R^{(6)}_{MN} R^{(6)MN} + R^{(6)}_{MNKL} R^{(6)MNKL} \right)\right] \right. \nonumber \\
&+&  \left. r_c^2 \int d^4 x \sqrt{-g^{(4)}} R^{(4)}
 \right\} + \int d^6 x {\cal L}_{bulk}
+ \int d^4 x {\cal L}_{brane}  \,.
\end{eqnarray}
Here $r_c=M_{4}/M_{6}^2$ is the induced gravity ``cross-over"
scale (marking the transition from 4D to 6D gravity),
 $M_{6}$ is the six-dimensional Planck mass
and  $M_{4}$ is the four-dimensional one.

 The metric as in the five-dimensional case is
\be ds_6^2=g_{\m\n}(r,\x)dx^\m
dx^\n+a^{2}(r,\x)d\x^2+L^2(r,\x)d\xi^2~,\label{6dmetric} \ee now
with $\mu=0,1,2,3$ whereas $\x,\xi$ denote the radial and angular
coordinates of the two extra dimensions (the $\x$ direction may
 or may not  be compact and the $\xi$ coordinate ranges form $0$ to $2\pi$).
 Note that we have assumed that there exists an azimuthal
symmetry in the system, so that both the induced four-dimensional
metric and the functions $a$ and $L$ do not depend on $\xi$.

The corresponding Einstein equations are
 \be
 G^{(6)N}_M + r_c^2
G^{(4)\n}_\m g_M^\m g^N_\n {\d(\x) \over 2 \pi L}-\alpha H_{M}^{N}
=\frac{1}{M^{4}_{6}} \left[-\L_{6}+T^{(B)N}_M+T^{(br)\n}_\m g_M^\m
g^N_\n {\d(\x) \over 2 \pi L}\right]~, \label{einsequat} \ee where
$H_{M}^{N}$ is the corresponding six-dimensional term of
(\ref{gaussbonnet})
 To obtain the braneworld equations we expand the
metric around the 3-brane as \be L(r,\x)=\b(r)\x+O(\x^{2})~, \ee
and as in the five-dimensional case the function $L$ behaves as
$L^{\prime}(r,0)=\beta(r)$, where a prime now denotes derivative
with respect to $\x$. The ``boundary" Einstein equations are \bea
G^{(4)}_{\m\n} \left(r_c^2+8\pi (1-\b)\a\right)|_0 &=& {1 \over
M_{6}^4}
T^{(br)}_{\m\n}|_0+ 2\pi (1-\b)g_{\m\n}|_0 \nonumber \\
&+& \pi L(r,\x)\,E_{\m\n}|_0-2\pi\b\a\,W_{\m\n}|_0
\label{einsteincomb}~, \eea where the term \be E_{\m\n}|_0 =
\left(K_{\m\n}-g_{\m\n}\,K\right)|_0~, \label{INDContrib} \ee
 appears because of the presence of the induced
gravity term in the gravitational action, while the term \bea
W_{\m\n}|_0 &=&
g^{\la\s}\partial_{\x}g_{\m\la}\partial_{\x}g_{\n\s}|_0-
g^{\la\s}\partial_{\x}g_{\la\s}\partial_{\x}g_{\m\n}|_0 \nonumber \\
&+&\frac{1}{2}g_{\m\n}\left[\left(g^{\la\s}\partial_{\x}g_{\la\s}\right)^2
-g^{\la\s}g^{\d\r}\partial_{\x}g_{\la\d}\partial_{\x}g_{\s\r}\right]\Big{|}_0~,
\label{GBContrib} \eea is the Weyl term due to the presence of the
Gauss-Bonnet term in the bulk~\cite{Bostock:2003cv}.
  The effective four-dimensional
 mass and cosmological constant are
\bea
 M^{2}_{Pl}&=&M_{6}^4 (r_c^2+8\pi (1-\b)\a)~,\\
\Lambda_{4}&=&\lambda-2\pi M_{6}^4 (1-\b)~, \label{La4} \eea where
$\lambda$ is the brane tension.

If we demand that the space in the vicinity of the conical
singularity is regular ($\de_\m \b=0$) then (\ref{einsteincomb})
simply becomes~\cite{Bostock:2003cv,Papantonopoulos:2005ma} \be
G^{(4)}_{\m\n} \left(r_c^2+8\pi (1-\b)\a\right)|_0 = {1 \over
M_{6}^4} T^{(br)}_{\m\n}|_0+ 2\pi (1-\b)g_{\m\n}|_0
\label{einsteincombsimple}~. \ee

\subsection{Black String-Like Solutions with pure conical singularity:
Case $\de_\m \b=0$}

In this subsection we make the assumption that the singularity is
purely conical. Thus, we will solve the bulk equations with a
constant deficit angle $\b$. We assume that the brane metric is
\be ds_{4}^{2}=-A(r)^{2}dt^{2}+A(r)^{-2}dr^{2}+r^{2}d\th^{2}
+r^{2}\,\sin^2\th\,d\phi^{2}~, \label{4dmetric}\ee where $0\leq r<
\infty$ is the radial coordinate, and $\phi$ has
 the usual periodicity $(0,2\pi)$.
 We will look for black string solutions of the
 Einstein equations~(\ref{einsequat}) using the
 six-dimensional metric~(\ref{6dmetric}) in the form
\be
ds_6^2=F^{2}(\x)\left(-A(r)^{2}dt^{2}+A(r)^{-2}dr^{2}+r^{2}d\th^{2}
+r^{2}\,\sin^2\th\,d\phi^{2}\right)+a^{2}(r,\x)d\x^2+L^2(r,\x)d\xi^2~.\label{6smetric}
\ee

 As we discussed in the previous section, the space outside the
 conical singularity is regular, therefore we  demand that the warp function $F(\chi)$ is
  also regular
 everywhere. We have split the general bulk energy momentum tensor $\tilde{T}_M^{(B)\,N}$ into
 a cosmological constant $\Lambda_{6}$ and a bulk energy momentum tensor $T_M^{(B)\,N}$.
 Moreover, we take $a(r,\chi)=1$. Then from the bulk equations
\be G^{(6)}_{MN}-\alpha
H_{MN}=\frac{1}{M^{4}_{6}}\,\left(-\Lambda_{6}\,g_{MN} +
T^{(B)}_{MN}\right)~, \label{eieq}\ee by taking the  combination
$rr - \th\th$ and $\x\x - \xi\xi$ of the Einstein equations we
respectively get \bea \left(\dot{A}^{2}+A
\ddot{A}-\frac{A^2}{r^2}+\fr{1}{r^2}\right)
\left[1-4\alpha \left(\frac{L''}{L}+\frac{F''}{F}+\frac{F'L'}{FL}\right)\right]=0~,\label{17} \\
\left(F''-\frac{F'L'}{L}\right)\left[1-\frac{2
\alpha}{F^{2}}\left(\dot{A}^{2}+A \ddot{A}+\fr{A^2}{r^2}+4\frac{A
\dot{A}}{r}-\fr{1}{r^2}+6F'^{2} \right)\right]=0\label{18}~. \eea
The $\x\x$ and $\xi\xi$ components of the Einstein equations are
\bea G_{\chi}^{\chi}&=&T_{\chi}^{\chi}~,\nonumber
\\G_{\xi}^{\xi}&=&T_{\xi}^{\xi}~, \eea and therefore to get (\ref{18})
we must take the difference $G_{\chi}^{\chi}-G_{\xi}^{\xi}=
T_{\chi}^{\chi}-T_{\xi}^{\xi}$,  and as the only remaining energy
contribution in the bulk is a cosmological constant,  the matter
in the extra two dimensions must satisfy the relation $T^{\x}_{\x}
= T^{\xi}_{\xi}$.

\subsubsection{Black String-Like Solutions of the Bulk Equations: Case 1}

We will consider first \be \dot{A}^{2}+A
\ddot{A}-\frac{A^2}{r^2}+\fr{1}{r^2}=0~, \ee which has as a
solution \be
A^{2}(r)=1+\frac{r^{2}}{L_4^{2}}-\frac{\z}{r}~,\label{BH?} \ee
where $L_4$ is the  length scale of the $AdS_{4}$ space, and
$\zeta$ is an integration constant. Then equation (\ref{18})
becomes \be
 \left(F''-\frac{F'L'}{L}\right)\left[1-12\frac{\alpha}{F^{2}}\left(\frac{1}{L_4^{2}}+F'^{2}
 \right)\right]=0\label{18a}~.
 \ee
From the above equation we have two cases:
\begin{itemize}
    \item \underline{Case 1a:} The first case is
\be F'^{2}-\frac{F^{2}}{12\alpha}+\frac{1}{L_4^{2}}=0~.
\label{1stcase} \ee This equation has the following solution
 \be
F(\x)=C_{1}\,e^{\frac{\x}{2\sqrt{3\alpha}}} +
C_{2}\,e^{\frac{-\x}{2\sqrt{3\alpha}}}~,\label{solu1}
 \ee where $
C_{1}$ and $C_{2}$ are integration constants which satisfy the
relation $C_{1}\,C_{2}=3\a/L_4^{2}$. The function $F(\x)$ is
regular and if we require on the position of the brane the
boundary condition $F^{2}(\x=0)=1$, the integration constants can
be expressed in terms of $\alpha$ and $L_4$ as
 \bea
C_{1}&=&\pm \frac{1 + \varepsilon \sqrt{1-12\frac{\a}{L_4^{2}}}}{2}~,  \nonumber\\
C_{2}&=&\pm \frac{1 - \varepsilon
\sqrt{1-12\frac{\a}{L_4^{2}}}}{2}~~,
 \eea
where $\varepsilon=\pm 1$ independently of the $\pm$ sign in
$C_{1}$ and $C_{2}$. Moreover, we need
$\partial_{\x}g_{\m\n}|_{(\x=0)}=0$, i.e.,
$\partial_{\x}F|_{(\x=0)}=0$, therefore $C_1=C_2=\frac{1}{2}$ and
$\a = \frac{L_4^{2}}{12}$. Thus, $F(\chi)= \cosh\left(
\frac{\chi}{L_4}\right)$. Substituting the above solutions into
the $tt$, $rr$, $\th\th$, and $\f\f$ components of the Einstein
equations we get a fine-tuned relation between $\alpha$ and
$\Lambda_{6}$ \be \Lambda_{6}=-\frac{5}{12
\alpha}=-\frac{5}{L_4^{2}}~.\label{tuning}
 \ee
Because of the positivity of $\alpha$ the six-dimensional bulk
space is Anti-de-Sitter. In addition we have a relation between
the six-dimensional cosmological constant $\La_6$ and the
$AdS_{4}$ length scale $L_4$. If we require the bulk equations to
have as a solution the Schwarzschild-AdS black hole (\ref{BH?})
with $\zeta \neq 0$ then consistency of the bulk equations requires
the $\x\x$ and $\xi\xi$ components of the energy-momentum tensor
to have the form \be T^{(B)\,\x}_{\x} = T^{(B)\,\xi}_{\xi} =
-\frac{6\, \a \, \z^2 \, }{r^6} \,
\frac{1}{F(\x)^4}~,\,\label{enermomx} \ee with the other
components to be zero. Notice that except for the warp function
$F(\chi)$ these components of the energy-momentum tensor do not
depend explicitly on $\chi$ but only on the radial coordinate on
the brane. We will give a detailed account of this dependence in
the next section.

    \item
\underline{Case 1b:} The second case is to consider
 \be
F''-\frac{F' L '}{L}=0~,\label{25}
 \ee which means that $L(\x)=L_0\,F'(\x)$.
In any subcase we recover in the same way the results of case 1a.
However $L(\x)$ is no more arbitrary and is given by
$L(\x)=\b\,L_4\,\sinh\left(\frac{\x}{L_4}\right)$, where we used
the boundary conditions $L(\x=0)=0$ and $L'(\x=0)=\b$.
\item
There are also two constant solutions for $F(\x)$ which read \bea
F(\x)&=& \pm 1, \\
A(r)&=&\sqrt{1+\frac{r^2}{L_4^2}-\frac{\z}{r}},
\\\label{nsolucte1} L(\x)&=&\b \,
\frac{\sinh{(\g\,\x)}}{\g},\\\label{bsolucte1}
\mathrm{with} \hspace{0.2in} \g &=& \frac{1}{L_4}\,\sqrt{\frac{1-\frac{L_4^2}{4\a}}{1-\frac{L_4^2}{12\a}}} \nonumber \\
\L_6&=& -\frac{6}{L_4^2} \left(1-\frac{2\a}{L_4^2}\right)\\
\mathrm{for} \hspace{0.2in} \z \neq 0 \hspace{0.2in}
T^{(B)\,\x}_{\x} &=& T^{(B)\,\xi}_{\xi} = -\frac{6\a \z^2}{r^6}.\,
\eea

The second one has the same $F(\chi)$ and $A(r)$ functions, as
well as $T^{(B)\,\x}_{\x} = T^{(B)\,\xi}_{\xi}$, but \bea
L(\x)&=&\b\,\x \, \frac{\sinh{\g}}{\g},\\\label{bsolucte2}
T^{(B)\,t}_{t} = T^{(B)\,r}_{r} = T^{(B)\,\th}_{\th} &=&
T^{(B)\,\f}_{\f} = \frac{3\left(4\a-L_4^2\right)}{L_4^4}. \eea

\end{itemize}

\subsubsection{Black String-Like Solutions of the Bulk Equations: Case 2}

In this section we chose from (\ref{17}) the equation  \be
1-4\alpha
\left(\frac{L''}{L}+\frac{F''}{F}+\frac{F'L'}{FL}\right)=0~.\label{case2}
\ee
\begin{itemize}
    \item
\underline{Case 2a:} The first one is to consider from (\ref{18})
 \be
F''-\frac{F' L '}{L}=0~,\label{case2a}
 \ee which means that $L(\x)=L_0\,F'(\x)$. Although this is not enough to solve (\ref{case2}), there is an exponential solution for both functions given by
\bea F(\x)&=&C_{1}\,e^{\frac{\x}{2\sqrt{3\alpha}}} +
C_{2}\,e^{\frac{-\x}{2\sqrt{3\alpha}}}~,\label{solu2} \\
L(\x)&=&L_0\,F'(\x) \,. \eea Substituting the above solution into
the $tt$, $rr$, $\th\th$, and $\f\f$ components of the Einstein
equations we get a fine-tuned relation between $\alpha$ and
$\Lambda_{6}$ \be \Lambda_{6}=-\frac{5}{12
\alpha}~.\label{tuning2} \ee If we choose as in the first case
$C_1C_2=\frac{3\a}{L_4^2}$ then we get a differential equation for
$A(r)$, which has the following solution \be
A^{2}(r)=1+\frac{r^{2}}{L_4^2}\pm \sqrt{1 + \frac{C_3}{L_4^2} +
\frac{C_4}{L_4^4}\,r} ~,\label{BH2?} \ee with the constraint $12\a
= L_4^2$ which relates the $6$-dimensional cosmological constant
$\La_6$ and the $AdS_{4}$ length scale $L_4$ as $\La_6 =
-\frac{5}{L_4^2}$, therefore imposing the boundary conditions
$F^{2}(\x=0)=1$ ($\partial_{\x}F|_{(\x=0)}=0$ is already
satisfied), $L(\x=0)=0$ and $L'(\x=0)=\b$ we have $F(\x) = \cosh
\left(\frac{\x}{2\sqrt{3\alpha}}\right)$ and $L(\x) =
2\,\sqrt{3\a}\,\b\, \sinh\left(\frac{\x}{2\sqrt{3\alpha}}\right)$
which satisfy all Einstein equations.
    \item
\underline{Case 2b:} The second case is to consider from
(\ref{18})
 \be
\left(F^{2}-12\a\,F'^{2}\right) -2\alpha\left(\dot{A}^{2}+A
\ddot{A}+\fr{A^2}{r^2}+4\frac{A \dot{A}}{r}-\fr{1}{r^2} \right)
=0~,\label{case2b} \ee The first term is a function of $\x$ while
the second one is a function of $r$. Therefore, each term should
be, in general, equal to a constant $\k$. We then have \bea
F^{2}-12\a\,F'^{2}&=&\k~, \\
2\alpha\left(\dot{A}^{2}+A \ddot{A}+\fr{A^2}{r^2}+4\frac{A
\dot{A}}{r}-\fr{1}{r^2} \right)&=&\k~, \eea which give \bea
F(\x)&=&C_{1}\,e^{\frac{\x}{2\sqrt{3\alpha}}} +
C_{2}\,e^{\frac{-\x}{2\sqrt{3\alpha}}}~, \\
A^{2}(r)&=&1+\frac{2\,C_3}{r^{2}} + \frac{C_4}{r} +
\frac{\k\,r^2}{12\,\a}, \eea with $C_1C_2 = \frac{\k}{4}$. No
solution can be found unless we set $\La_6=-\frac{5}{12\a}$. Then
we need to solve the following differential equation \be -L(\x) +
\sqrt{3\a}\left(\frac{1-\frac{\k}{4\,C_1^2}\,e^{\frac{-\x}{\sqrt{3\a}}}}
{1+\frac{\k}{4\,C_1^2}\,e^{\frac{-\x}{\sqrt{3\a}}}}\right)\,
L'(\x) + 6\a\,L''(\x) = 0, \label{diff2a} \ee which has the
following solution \be L(\x) =
\frac{4\,C_1^2\,C_5\,e^{\frac{\x}{\sqrt{3\alpha}}}}{\k}\,
\,
_2F_1\left[\frac{1}{2},2,\frac{5}{2},-\frac{4\,C_1^2}{\k}\,e^{\frac{\x}{\sqrt{3\alpha}}}\right]\,,
\ee being $_2F_1$ the hypergeometric function of the second kind.
The $\x\x$ and $\xi\xi$ components of the Einstein equations
impose us $C_1=0$, but then we cannot have
$\partial_{\x}\,g_{\m\n} = 0$. Therefore we must have $C_1\neq0$
and we have to consider a specific $r$ and $\x$ dependent $\x\x$
and $\xi\xi$ components of the energy momentum tensor given by \be
T^{(B)\,\x}_{\x} = T^{(B)\,\xi}_{\xi} = -\frac{2\a \left(40\,
C_3^2 + 24\,C_3\,C_4\,r + 3\,C_4^2\,r^2 \right)}{r^8}\,
\frac{1}{F(\x)^4}.\, \ee Imposing the boundary conditions
$F^{2}(\x=0)=1$, $L(\x=0)=0$ and $L'(\x=0)=\b$ we get \bea
C_{1}&=&\pm \frac{1 + \varepsilon \sqrt{1-\k}}{2}~, \\
C_{2}&=&\pm \frac{1 - \varepsilon \sqrt{1-\k}}{2}~, \\
\k &=& \b, \\
C_{5} &=& \frac{5\,\sqrt{3\a}\,\b^3}{\eta^2}\,\Big{(}5\,\b\,
\, _2F_1\left[\frac{1}{2},2,\frac{5}{2},-\frac{\eta^2}{\b}\right]\, \nonumber \\
&-&2\,\eta^2\,
\, _2F_1\left[\frac{3}{2},3,\frac{7}{2},-\frac{\eta^2}{\b}\right]\,\Big{)}, \\
\mathrm{with} \hspace{0.2in} \eta &=& 1+\sqrt{1-\b}. \nonumber
\eea Moreover, we must have $\partial_{\x}F|_{(\x=0)}=0$ therefore
$C_1=C_2=\frac{1}{2}$ i.e. $\k=1$ which imposes $C_5=0$. Therefore
there is no solution.
\item
There is also a constant solution for $F(\x)$ which gives \bea
F(\x)&=& \pm 1, \label{consolu}\\
A(r)^2&=&1+\frac{r^2}{4\a}-\frac{\sqrt{3}}{12\a}
\sqrt{2\,r^4 - 3\,C_4\,r +48\,\a\,\left(\a - C_3\right)}, \label{nsolucte2}\\
L(\x)&=&2\sqrt{\a}\,\b \, \sinh{\frac{\x}{2\sqrt{\a}}},\label{bsolucte2a} \\
\L_6&=& -\frac{1}{4\a}.\, \label{consolu2}\eea

\end{itemize}

\subsubsection{Localization of the Bulk Black Hole on the Brane}

In order to complete our solution with the introduction of the
brane we must  solve the corresponding junction conditions given
by the Einstein equations on the brane (\ref{einsteincombsimple})
using the induced metric on the brane given by (\ref{4dmetric}).

Equation (\ref{einsteincombsimple}) can be written as \be
\frac{T^{(br)\,\n}_{\m}|_0}{M^6_4} = \left(r_c^2 +
8\pi\,\left(1-\b\right)\,\a\right)\,G^{(4)\,\n}_{\m}|_0
-2\pi\,\left(1-\b\right)\,g^{\,\,\n}_{\m}|_0, \label{Tmumu} \ee
Moreover, the $(\x\x)$ component of the six-dimensional Einstein
tensor evaluated at $\chi = 0$ is \be -\frac{1}{2}\,R^{(4)}|_0 -
\frac{\a}{2} \left( R^{(4)\,2} - 4R^{(4)\,2}_{\m\n}
+R^{(4)\,2}_{\m\n\k\la} \right)\Big{|}_0 =
\frac{1}{M^4_6}\,T^{(B)\,\chi}_{\chi}|_0
-\frac{\La_6}{M^4_6}|_0,\, \label{6DBulkrr} \ee which gives the
form of the $(\x\x)$ component of the bulk energy momentum tensor
in terms of brane quantities \be
\frac{1}{M^4_6}\,T^{(B)\,\chi}_{\chi} |_0 =
-\frac{1}{2}\,R^{(4)}|_0 - \frac{\a}{2} \left(R^{(4)\,2} -
4R^{(4)\,2}_{\m\n} +R^{(4)\,2}_{\m\n\k\la} \right)\Big{|}_0
+\frac{\La_6}{M^4_6}|_0. \, \label{Txx} \ee

\begin{itemize}
    \item \underline{For case 1a we have:}
\bea
A^2(r) &=& 1 + \frac{r^2}{L_4^2} - \frac{\z}{r}, \nonumber\\
F(\x) &=& \cosh\left({\frac{\x}{2\sqrt{3\alpha}}}\right),
\nonumber \eea In this case $L(\x)$ is arbitrary, and we have the
constraint $\a = \frac{L_4^2}{12}$. In addition, \bea
\Lambda_{6}&=&-\frac{5}{12 \alpha}=-\frac{5}{L_4^2}, \label{C1,2}\\
T^{(B)\,\x}_{\x} = T^{(B)\,\xi}_{\xi} &=& -\frac{6\, \a \, \z^2 \,
}{r^6} \, \frac{1}{F(\x)^4}.\, \label{Txxold} \eea Then
(\ref{Txx}) is consistent with (\ref{Txxold}) whereas
(\ref{Tmumu}) gives \be \frac{T^{\n}_{\m}}{M^4_6} =
3\frac{r_c^2}{L_4^2}\,\d^{\n}_{\m}. \ee
    \item \underline{For case 1b we have:}
\bea
A^2(r) &=& 1 + \frac{r^2}{L_4^2} - \frac{\z}{r}, \nonumber\\
F(\x) &=& \cosh \left(\frac{\x}{L_4} \right), \label{Fsol1b}\\
L(\x) &=& \b\,L_4\,\sinh\left(\frac{\x}{L_4}\right),
\label{Lsol1b} \eea with the constraint $\a = \frac{L_4^2}{12}$
and \bea
\La_6 &=& -\frac{5}{12\a} = -\frac{5}{L_4^2}, \nonumber\\
T^{(B)\,\x}_{\x} = T^{(B)\,\xi}_{\xi} &=& -\frac{6\, \a \, \z^2 \,
}{r^6} \, \frac{1}{F(\x)^4}. \label{Txxold1}\, \eea From
(\ref{Tmumu}) we get $\frac{T^{\n}_{\m}}{M^4_6} =
3\frac{r_c^2}{L_4^2}\,\d^{\n}_{\m}$ whereas (\ref{Txx}) and
(\ref{Txxold1}) are consistent.
    \item \underline{For case 2a we have:}
\bea
A^{2}(r)&=&1+\frac{r^{2}}{L_4^{2}}\pm \sqrt{1 + \frac{C_3}{L_4^2} + \frac{C_4}{L_4^4}\,r}, \nonumber\\
F(\x) &=& \cosh \left(\frac{\x}{2\sqrt{3\alpha}}\right), \nonumber\\
L(\x) &=& 2\,\sqrt{3\a}\,\b\, \sinh\left(\frac{\x}{2\sqrt{3\alpha}}\right), \nonumber\\
\La_6 &=& -\frac{5}{12\a} = -\frac{5}{L_4^2}, \nonumber \eea In
this case (\ref{Txx}) and (\ref{Tmumu}) give respectively some
complicated $r$ dependent expressions for $T^{\x}_{\x}$ and
$T^{\m}_{\m}$, as well as for the solution
(\ref{consolu})-(\ref{consolu2}).

    \item \underline{For case 2b we have no solution.}
\end{itemize}
Our results are summarized in Table \ref{table2}.

\begin{table}[here]
\begin{center}
\begin{tabular}{|c|c|c|c|c|}
\hline $A^2(r)$ & $F(\chi)$ & $L(\chi)$ & $-\Lambda_6$ &
Constraints \& $T^{(B)}$\\ \hline
$1+\frac{r^2}{L_4^2}-\frac{\zeta}{r}$ & $\cosh\left( \frac{\chi}{2\sqrt{3\alpha}}\right)$ & $\forall L(\chi)$  & $\frac{5}{12\a}$ & $\a=\frac{L_4^2}{12}$, \\
&&&&$T^{\chi}_\chi = T^{\xi}_\xi = - \frac{6\a \zeta^2}{r^6 F(\chi)^4}$\\
$1+\frac{r^2}{L_4^2}-\frac{\zeta}{r}$ & $\cosh\left( \frac{\chi}{2\sqrt{3\alpha}}\right)$ & $2\sqrt{3\alpha}\beta \sinh\left( \frac{\chi}{2\sqrt{3\alpha}}\right)$ & $\frac{5}{12\a}$ & $\a=\frac{L_4^2}{12}$, \\
&&&&$T^{\chi}_\chi = T^{\xi}_\xi = - \frac{6\a \zeta^2}{r^6 F(\chi)^4}$ \\
$1+\frac{r^2}{L_4^2}-\frac{\zeta}{r}$ & $\pm 1$ & $\frac{\b}{\g} \, \sinh{(\g\,\x)}$ & $\frac{6}{L_4^2} \left(1-\frac{2\a}{L_4^2}\right)$ & $\g = \frac{1}{L_4}\,\sqrt{\frac{1-\frac{L_4^2}{4\a}}{1-\frac{L_4^2}{12\a}}}$, \\
&&&&$T^{\chi}_\chi = T^{\xi}_\xi = - \frac{6\a \zeta^2}{r^6}$\\
$1+\frac{r^2}{L_4^2}-\frac{\zeta}{r}$ & $\pm 1$ & $\frac{\b}{\g} \chi \, \sinh{\g}$ & $\frac{6}{L_4^2} \left(1-\frac{2\a}{L_4^2}\right)$ & $\g = \frac{1}{L_4}\,\sqrt{\frac{1-\frac{L_4^2}{4\a}}{1-\frac{L_4^2}{12\a}}}$, \\
&&&&$T^{\chi}_\chi = T^{\xi}_\xi = - \frac{6\a \zeta^2}{r^6}$, \\
&&&&$T_t^{t}=T_r^{r}=T_\theta^{\theta}=T_\phi^{\phi}= \frac{3(4\a-L_4^2)}{L_4^2}$\\
(\ref{BH2?})& $\cosh \left(\frac{\x}{2\sqrt{3\alpha}}\right)$ & $2\,\sqrt{3\a}\,\b\, \sinh\left(\frac{\x}{2\sqrt{3\alpha}}\right)$ & $\frac{5}{12\a}$ & $\a=\frac{L_4^2}{12}$ \\
(\ref{nsolucte2}) &
$\pm 1$ & $2\sqrt{\a}\,\b \, \sinh{\left(\frac{\x}{2\sqrt{\a}}\right)}$ & $\frac{1}{4\a}$ & $\a=\frac{L_4^2}{4}$ \\
\hline
\end{tabular}
\end{center}
\caption{Black String-Like Solutions in Six-Dimensional
Braneworlds
of Codimension-2}\label{table2}
\end{table}

\subsection{Curvature singularity: Case $\de_\m \b\neq0$}

In this section we relax the assumption of the purely conical
singularity. Therefore, in general $\partial_{\x}\,g_{\m\n}\neq0$
and $\b(r)$ is a function of r.

\subsubsection{Black String-Like Solutions of the Bulk Equations}

In this case the combination of the $rr - tt$ components of the
bulk equations (\ref{eieq}) give \be rr - tt \hbox{: }
-\frac{A^2\,\ddot{L}}{r^2\,F^4\,L}\,\left[4\a-4\a
A^2+r^2\left(F^2-4\a F'^2-8\a FF''\right)\right]. \label{rr-tt}
\ee If we want to keep this factorized form, we can choose
$\ddot{L}=0$ which will not simplify our task. Therefore, we
consider in general $\ddot{L}\neq0$. The other possibility is to
consider the term in square brackets equal to zero.

\begin{itemize}
\item \underline{Case 1:}
In this case the term in square brackets of (\ref{rr-tt}) is equal
to zero. Thus we will have $T^{(B)\,r}_r = T^{(B)\,t}_t$ and we
need to solve the following equations \bea
4l^2 \a - 4\a A^2 &=& -\k r^2, \\
F^2-4\a F'^2-8\a FF'' &=& \k~, \eea where $\k$ is a constant. Then
the solutions are \bea
A^2(r)&=&1 + \frac{\k}{4\a}\,r^2, \\
F(\x)&=&C_1\,e^{\frac{\x}{2\sqrt{3\a}}}+C_2\,e^{\frac{-\x}{2\sqrt{3\a}}},
\eea with $\k=\frac{4\,C_1\,C_2}{3}$. If we redefine
$\frac{4\a}{\k}=L_4^2$ we get \bea
A^2(r)&=&1+\frac{r^2}{L_4^2}~, \\
F(\x)&=&C_1\,e^{\frac{\x}{2\sqrt{3\a}}}+C_2\,e^{\frac{-\x}{2\sqrt{3\a}}}
\,\,\,\, \mathrm{with} \,\,\,\, C_1\,C_2 = \frac{3\a}{L_4^2}~.
\eea Then all the bulk Einstein equations are satisfied for
$\L_{(6)} = -\frac{5}{12\a}$ and with no matter in the bulk.
Furthermore, if we require on the position of the brane the
boundary condition $F^{2}(\x=0)=1$, the integration constants can
be expressed as in the case $1a$ with constant deficit angle, in
terms of $\alpha$ and $L_4$
 \bea
C_{1}&=&\pm \frac{1 + \varepsilon \sqrt{1-12\frac{\a}{L_4^{2}}}}{2}~,  \nonumber\\
C_{2}&=&\pm \frac{1 - \varepsilon
\sqrt{1-12\frac{\a}{L_4^{2}}}}{2}~~, \label{C1C2new}
 \eea
where $\varepsilon=\pm 1$ independently of the $\pm$ sign in
$C_{1}$ and $C_{2}$. Moreover, $L(r,\x)$ is arbitrary.

\item \underline{Case 2:}
In this case the factorized equation (\ref{rr-tt}) is not equal to
zero (i.e. $T^{(B)\,r}_r \neq T^{(B)\,t}_t$) but then the bulk
Einstein equations cannot be solved in general. However, if we
consider that $A(r)$ has the same form as in the previous
subsection  for cases $1a$ and $1b$, \be
A^2(r)=1+\frac{r^2}{L_4^2}-\frac{\z}{r}~, \ee then the combination
$\th\th - tt$ and $\x\x - \xi\xi$ of the bulk equations
(\ref{eieq}) can respectively be factorized as \bea \th\th - tt
&\hbox{: }& \frac{\left(2l^2r-3\z\right) \dot L \left[8\a \z
L_4^2+4\a r^3+r^3 L_4^2\left(4\a F'^2 + 8\a
FF''-F^2\right)\right]}{2r^5 L_4^2 F^4 L} \nonumber \\ &&-
\frac{12\alpha\zeta r \ddot L \,(r^3 + l^2L_4 ^2 r -L_4^2\zeta)}
{2r^5 L_4^2 F^4 L}~, \hspace{0.5in}\\
\x\x - \xi\xi &\hbox{: }& \left(12\a-L_4^2 F^2+12\a L_4^2
F'^2\right) \times \nonumber \eea \be \times
\frac{\left[\dot{L}\left(4r^3+2 l^2 L_4^2 r - \z L_4^2\right) +
r\,\ddot{L}\,\left(r^3+L_4^2\,r\,l^2-\z\,L_4^2\right) + 4r^2 L_4^2
F \left(F'L'- F''L\right)\right]} {r^2 L_4^4 F^4 L}~,
\hspace{-0.12in}\label{xixixx} \ee where a  dot denotes
derivatives with respect to $r$. We note here that in the $\th\th
- tt$ combination the first term in brackets can never be zero
while the second one cannot be solved analytically. Therefore here
$T^{r}_{r} \neq T^{\th}_{\th}$. In the $\x\x - \xi\xi$ combination
the second term in brackets can not also be solved analytically.
Therefore the only term which can be equal to zero in order to
keep a factorized form is the first bracket in (\ref{xixixx}).
 Then we get that
$F(\x)=C_1\,e^{\frac{\x}{2\sqrt{3\a}}}+C_2\,e^{\frac{-\x}{2\sqrt{3\a}}}$
with $C_1\,C_2=\frac{3\a}{L_4^2}$. Finally, the bulk Einstein
equations are satisfied for $\L_{(6)}=-\frac{5}{12\a}$ and for
$\z=0$, unless we have the following bulk energy momentum tensor
\bea
T^{(B)\,t}_t &=& -T^{(B)\,\th}_{\th} = -T^{(B)\,\f}_{\f} = \frac{2\a \z}{r^5 L F^4} \nonumber \\
&&\times \left[\left(3\z-2l^2r\right)\dot{L}
+2r^2\ddot{L}\left(l^2+\frac{r^2}{L_4^2}-\frac{\z}{r}\right)\right], \\
T^{(B)\,r}_r &=& \frac{2\a \z}{r^5 L F^4}\left(3\z-2l^2r\right)\dot{L}~, \\
T^{(B)\,\x}_{\x} &=& T^{(B)\,\xi}_{\xi} = -\frac{6\, \a \, \z^2 \,
}{r^6} \, \frac{1}{F(\x)^4}~. \eea Furthermore, if we require on
the position of the brane the boundary condition $F^{2}(\x=0)=1$,
the integration constants can be expressed as in (\ref{C1C2new}).
Moreover, $L(r,\x)$ is arbitrary.
\end{itemize}

\subsubsection{Localization of the Bulk Black Hole on the Brane}

In order to complete our solution with the introduction of the
brane we must  solve the corresponding junction conditions given
by the Einstein equations on the brane (\ref{einsteincomb}) using
the induced metric on the brane given by (\ref{4dmetric}).
Equation (\ref{einsteincomb}) can be written as \bea
\frac{T^{(br)\,\n}_{\m}|_0}{M^6_4} &=& \left(r_c^2+8\pi
(1-\b)\a\right)\,G^{(4)\,\m}_{\n}|_0
- 2\pi (1-\b)\d^{\m}_{\n}|_0 \nonumber \\
&-& \pi L(r,\x)\,E^{\m}_{\n}|_0+2\pi\b\a\,W^{\m}_{\n}|_0.
\label{Tmneinsteincomb(r)}~, \eea Moreover, the $(\x\x)$ component
of the six-dimensional Einstein tensor evaluated at $\chi = 0$ is
given in terms of brane quantities as \bea
\frac{1}{M^4_6}\,T^{(B)\,\chi}_{\chi} \Big{|}_0 &=&
-\frac{1}{2}\,R^{(4)}\Big{|}_0 - \frac{\a}{2} \left(R^{(4)\,2} -
4R^{(4)\,2}_{\m\n}
+R^{(4)\,2}_{\m\n\k\la} \right)\Big{|}_0 \nonumber\\
&-& \frac{K'}{4}\Big{|}_0 - \frac{1}{8}K^{\n}_{\m}K^{\m}_{\n}|_0
+\frac{g'L'}{4gL}\Big{|}_0 +
\frac{\bigtriangledown^{(4)}_{\m}\partial^{\m}L}{L}\Big{|}_0
+\frac{\La_6}{M^4_6}\Big{|}_0. \, \label{Txx(r)} \eea In the above
relation we have  \bea
\frac{K'}{4}\Big{|}_0 &=& 2\left(\frac{F''}{F}-\frac{F'^2}{F^2}\right)\Big{|}_0 \label{IND1}\\
\frac{1}{8}K^{\n}_{\m}K^{\m}_{\n}|_0 &=& 2\frac{F'^2}{F^2}\Big{|}_0, \label{IND2}\\
\frac{g'L'}{4gL}\Big{|}_0 &=& \frac{F'L'}{2FL}\Big{|}_0, \label{pb}\label{IND3}\\
\frac{\bigtriangledown^{(4)}_{\m}\partial^{\m}L}{L}\Big{|}_0 &=&
\frac{1}{F^2\,L}\left[2\dot{L}\left(\frac{A^2}{r}+A\dot{A}\right)+A^2\ddot{L}\right]\Big{|}_0,\label{IND4}
\eea and we can see that requiring $L(r,\x=0)=0$ all terms are
regular except (\ref{pb}) which has a $\frac{1}{\x}$ contribution
which is singular. This can only be avoided if $F'|_0=0$, thus $\a
= \frac{L_4^2}{12}$, i.e., $\b$ constant. Another way to make the
relation (\ref{Txx(r)}) regular is to take the pure Gauss-Bonnet
case, where we do not take under consideration the induced gravity
term in the action. In this case the bulk solutions are the same
and we do not have the contributions (\ref{IND1})-(\ref{IND4}) in
(\ref{Txx(r)}) which becomes as (\ref{Txx}). Then for both cases 1
and 2 if we want to match the $T^{(B)\,\x}_{\x}$ component of the
bulk solution with the one derived in (\ref{Txx(r)}) we must have
the relation $\a = \frac{\la^2}{12}$ which gives the constant $\b$
case. Thus, the relation (\ref{Txx(r)}) between bulk and brane
quantities in order  to be regular in the vicinity of the conical
singularity requires the deficit angle to be constant.

\section{The r$\hat{o}$le of the Gauss-Bonnet Term}

It is known that  from a Ricci flat (D-1)-dimensional solution
 a D-dimensional solution can be generated which satisfies the
Ricci flat D-dimensional Einstein equations~\cite{Brecher:1999xf}.
This procedure can also be applied if there is a D-dimensional
negative cosmological constant in the bulk. This result was used
in~\cite{Chamblin:1999by} to construct the five-dimensional black
string in codimension-1 branes.

If there is a Gauss-Bonnet term in the bulk there is a drastic
change in this result~\cite{Barcelo:2002wz,Kobayashi:2004hq}. In
the five-dimensional case consistency of the four-dimensional
Einstein equations forces the four-dimensional Gauss-Bonnet term
projected on the brane to be constant~\cite{Barcelo:2002wz}
\footnote{A similar relation obtained in~\cite{Barcelo:2002wz}
involving the Gauss-Bonnet term was presented
in~\cite{Molina:2008kh} in a different context.}. This implies
that there could not exist black string solutions of the
type in~\cite{Chamblin:1999by} with a Gauss-Bonnet term in the bulk
in codimension-1 braneworlds.

In codimension-2 braneworlds there is a relation connecting the
Gauss-Bonnet term projected on the brane with the components of
the bulk energy-momentum tensor corresponding to the extra
dimensions~\cite{Papantonopoulos:2005ma}. In six dimensions it
reads \be -\frac{1}{2}\,R^{(4)}|_0 - \frac{1}{2} \a  \left(
R^{(4)\,2} - 4R^{(4)\,2}_{\m\n} +R^{(4)\,2}_{\m\n\k\la}
\right)\Big{|}_0 = \frac{1}{M^4_6}\,T^{(B)\,\chi}_{\chi}|_0
-\frac{\La_6}{M^4_6}|_0\,. \label{6DBulkrr1} \ee All bulk
solutions have to satisfy this relation which acts as a
consistency relation. In spite of the fact that in four dimensions the
Gauss-Bonnet term is a topological invariant, when it is projected
on the brane, it leaves its traces through this relation. For the
Schwarzschild-AdS solution of the form (\ref{BH?}) the square of
the Riemann tensor reads \be
 R_{\mu\nu\kappa\lambda}^2=\frac{192\zeta^{2}e^{\frac{4\chi}{L_4}}}{(1+e^{\frac{2\chi}{L_4}})^{4}r^{6}}
+\frac{60}{L_4^{4}}~,\label{riemanntensor} \ee while the Ricci
scalar and Ricci tensor are constants. Therefore, for the relation
(\ref{6DBulkrr1}) to be satisfied the bulk energy-momentum tensor
$T^{(B)\,\chi}_{\chi}|_0$ has to scale as $1/r^{6}$ with the right
coefficients. This is actually what happens considering the result
(\ref{enermomx}). Moreover, it is easy to verify that
relation~(\ref{6DBulkrr1}) is satisfied substituting the relevant
quantities. Thus, the presence of the Gauss-Bonnet term in the
bulk, which acts as a source term because of its divergenceless
nature, dictates the form of matter that must be introduced in the
bulk in order to sustain a black hole on the brane\footnote{Black
hole solutions in codimension-2 braneworlds were also recently
discussed in~\cite{Charmousis:2008bt}.}.

For the physically, most  interesting  solutions of the
Schwarzschild-AdS black hole on the brane, we found that there
must be non-trivial matter in the extra two dimensions given by
(\ref{enermomx}). These components of the energy-momentum tensor
depend on the radial distance on the brane $r$ and on one of the
extra dimensions $\chi$ through the warp function $F(\chi)$.
Therefore, if we go far away from the brane (large $\chi$) because
of the form of the warp function (see Table 2) the energy momentum
tensor coming from the bulk decouples. This means that on the
brane we have standard four-dimensional gravity without any
corrections from the bulk. On the contrary, near the brane the $1/
r^{6}$ term dominates (the warp function goes to a constant)
giving a strong modification of the four-dimensional gravity on
the brane.

In five-dimensions a similar relation to~(\ref{6DBulkrr1}) holds.
Then, if we use the BTZ solution of Table 1 of section 2, the
corresponding relation in five-dimensions is automatically
satisfied. The reason is that the BTZ black hole does not have an
$r=0$ curvature singularity~\cite{Banados:1992gq} and, therefore,
all the curvature invariants appearing in the relation are
constants. Also the BTZ solution does not require matter in the
bulk~\cite{CuadrosMelgar:2007jx}. Thus, the corresponding relation
to~(\ref{6DBulkrr1}) in  five dimensions is trivially satisfied,
allowing the existence of a black string-like solution in
five-dimensional braneworlds of codimensionality two.

The situation is more subtle for the sort distance BTZ-corrected
solution of section 2. This black hole has $1/r$ curvature
singularity giving, therefore, a non-constant Krestschmann scalar
proportional to $1/r^{6}$. This implies that for the relation to
hold the combination of the three-dimensional squared Ricci scalar
and the squared Ricci tensor  should also be proportional to
$1/r^{6}$ with the appropriate coefficients. These curvature
invariants can be obtained solving the three-dimensional Einstein
equations on the brane~(\ref{einsteincomb3}). In order to get a
non-trivial solution matter should be introduced on the brane, and
this is actually what happens as it was shown
in~\cite{CuadrosMelgar:2007jx} (see relation
(\ref{braneEnerMom})).

In the five-dimensional case we have found that the matter
necessary to sustain the BTZ-corrected black hole solution on the
brane is provided by a scalar field conformally coupled to the BTZ
black hole. In six dimensions it is not clear to what system the
"holographic matter" necessary to sustain the Schwarzschild-AdS
black hole on the brane, corresponds. Considering the similarities
between the five and six-dimensional cases it might correspond
also to a scalar field coupled to the six-dimensional
gravitational action.

\section{Conclusions}
 We discussed black hole localization on an infinitely thin
3-brane of codimension-2 and its extension into a six-dimensional
AdS bulk. To have a four-dimensional gravity on the brane we
introduced a six-dimensional Gauss-Bonnet term in the bulk and an
induced gravity term on the brane. We showed that  a
Schwarzschild-AdS black hole can be localized on the brane which
is extended into the bulk with a warp function.
 Consistency of the six-dimensional bulk
equations requires a fine-tuned relation between the Gauss-Bonnet
coupling constant and the length of the six-dimensional AdS space.
The use of this fine-tuning gives to the non-singular horizon the
shape of a throat up to the horizon of the AdS space with no other
curvature singularities except the Schwarzschild string-like
singularity.

If the deficit angle is constant, independent of the radial
coordinate of the brane, there is a consistency relation between
the Gauss-Bonnet term projected on the brane and the
energy-momentum tensor of the two extra dimensions. This relation
for the Schwarzschild-AdS black hole solution on the brane
requires the presence of a form of "holographic matter" in the
extra dimensions which scales as $1/r^{6}$. This gives a strong
modification of gravity at short distances while standard GR is
obtained only at large distances.

If the deficit angle is variable,  the effective four-dimensional
Einstein equations on the brane acquire  extra terms  related to
the projection of the Weyl tensor on the brane. Also, the
constraint relation connecting the Gauss-Bonnet term projected on
the brane and the bulk energy-momentum tensor is more
involved, and in spite of the fact that the Schwarzschild-AdS black hole
solution on the brane is still a solution of the bulk equations,
it gives an inconsistency forcing the deficit angle to be
constant.

The presence of the  Gauss-Bonnet term is important in our
considerations. It allows the existence of black string solutions
in five-dimensions and in six dimensions it specifies the form of
matter which is needed  in the bulk in order to sustain a black
hole on the brane. It would have been interesting to find out what
modifications the gravitational action is needed in order to
obtain bulk solutions without the need of matter in the extra
dimensions.

\section*{Acknowledgments}

We benefited from the discussions, comments, and remarks we had
with Christos Charmousis, Roy Maartens, Antonis Papazoglou, and
Ricardo Troncoso. This work was supported by the NTUA research
program PEVE07. The work of B.C-M. is supported by Fondo Nacional
de Desarrollo Cient\'{i}fico y Tecnol\'ogico (FONDECYT), Chile,
under grant 3070009.

\end{document}